\documentclass{ws-p8-50x6-00}

\newcommand{\beq}{\begin{equation}}
\newcommand{\eeq}[1]{\label{#1} \end{equation}}

\begin{document}

\title{\null\vspace{-1cm}\hfill {\bf KFKI - 1999 - 06/A} \\[2.5ex]
Signal of Partial $U_A(1)$ Symmetry Restoration from Two-Pion
Bose-Einstein Correlations 
}

\author{
T. Cs\"{o}rg\H{o}
}

\address{
MTA KFKI RMKI, H-1525 Budapest 144, P.O. Box 49, Hungary
\\E-mail:
 csorgo@sunserv.kfki.hu
}

\author{
D. Kharzeev
}

\address{
RIKEN-BNL Research Center, Brookhaven National Laboratory, 
Upton, NY 11973
\\
E-mail: kharzeev@bnl.gov
}

\author{
S.E. Vance 
}

\address{
Brookhaven National Laboratory\\
E-mail: vance@bnl.gov
}  

\date{October 1, 1999}

\maketitle

\abstracts{
The intercept parameter of the two-pion Bose-Einstein correlation
functions at low $p_t$ is shown to play the role of an 
{\it experimentally accessible measure}
of the (partial) restoration of $U_A(1)$ symmetry 
in ultra-relativistic nuclear collisions.
}

\section{$U_A(1)$ symmetry restoration and the core-halo model}
In this conference contribution, we follow the lines of
ref.~\cite{vcsk} to show that a link exists between the 
symmetry properties of hot and dense hadronic matter and
the strength of the two-pion Bose-Einstein correlation functions.
 
In the chiral limit ($m_u = m_d = m_s = 0$), QCD possesses a $U(3)$ chiral 
symmetry. When broken spontaneously, $U(3)$ implies the existence of  
nine massless Goldstone bosons.  In nature, 
however, there are only eight light pseudoscalar mesons, 
a discrepancy which is resolved by the Adler-Bell-Jackiw $U_A(1)$ anomaly; 
the ninth would-be Goldstone boson gets a mass as a consequence of the 
nonzero density of topological charges in the QCD vacuum
\cite{witten,veneziano}.  
In recent papers \cite{kapusta96etap,xnwang}, 
it was argued that the partial restoration of $U_A(1)$ symmetry of QCD 
and related decrease of the $\eta'$ mass 
\cite{pisarski,kunihiro,shuryak,hatsuda} 
in regions of sufficiently hot and dense 
matter should manifest itself in a factor of 3 to 50 
increase~\cite{kapusta96etap} in the production of $\eta'$ mesons,  
relative to nuclear interactions which do not produce the phase 
transition. 

It was also observed, however, that the $\eta'$ decays are characterized 
by a small signal-to-background ratio in the direct two-photon decay mode. 
We have shown in ref.~\cite{vcsk},
that the momentum dependence of $\lambda_*$, that characterizes the 
strength of  Bose-Einstein correlations of pions, 
provides an experimentally
well observable signal for partial $U_A(1)$ restoration.

As was shown in several papers\cite{csorgo_96ch,nickerson_97}, 
at incident beam energies of 200 AGeV 
at the CERN SPS, the space-time structure of pion emission in high 
energy nucleus-nucleus collisions can be separated into two regions: 
the {\it core} and the {\it halo}.
The pions which are emitted from the core or central region
are either 
produced from a direct production 
mechanism such as the hadronization of wounded string-like 
nucleons in the collision region, 
rescattering as they flow outward with a rescattering time 
on the order of 1 fm/c, or  they are produced from  
the decays of short-lived hadronic resonances such as the $\rho$, $N^*$,
$\Delta$  and $K^*$, whose decay time is also on the order of 1-2 fm/c.  
This core region is resolvable by Bose-Einstein correlation (BEC) 
measurements.  The halo region, however, 
consists of the decay products of long-lived hadronic resonances 
such as the $\omega$, $\eta$, $\eta '$ and $K^0_S$ whose 
lifetime is greater than 20 fm/c.  This halo region is not resolvable by 
BEC measurements but it contributes to the reduction of 
the effective intercept parameter, $\lambda_*$.

The two-particle Bose-Einstein correlation function is defined as
\beq 
C(\Delta k, K) = \frac{N_2({\bf p_1, p_2})}{N_1({\bf p_1}) N_1({\bf p_2})}, 
\eeq{def_becf}
where the inclusive $1$ and $2$-particle 
invariant momentum distributions  are 
\beq 
N_1({\bf p}_1) = \frac{1}{\sigma_{in}} E_1 \frac{d\sigma}{ d{\bf p}_1 },
\qquad
N_2({\bf p}_1, {\bf p}_2) = 
\frac{1}{\sigma_{in}} E_1 E_2 \frac{d\sigma}{ d{\bf p}_1 d{\bf p}_2} ,
\eeq{lpart}
with $p = (E_{\bf p},{\bf p})$, 
$\Delta k = p_1 - p_2$, and  $K = {(p_1 + p_2)/}{2}$.

From the four assumptions made in the core-halo model\cite{csorgo_ch97}, 
the Bose-Einstein correlation function is found to be
\beq 
C(\Delta k,K) \simeq  1 + \lambda_* R_c(\Delta k,K),
\eeq{becf2}
where the effective intercept parameter $\lambda_*$
and the correlator of the core, $R_c(\Delta k, K)$ are defined, 
respectively, as
\beq 
\lambda_* = \lambda_*(K = p; Q_{min}) = \left [\frac{N_c({\bf p})}
{N_c({\bf p})+N_h({\bf p})}\right ]^2
\eeq{lambda1}
and
\beq 
R_c(\Delta k,K) = \frac{| \tilde{S}_c(\Delta k,K)|^2}{|
\tilde{S}_c(\Delta k = 0,K = p)|^2} \; .
\eeq{correlator}
Here, $\tilde{S}_c(\Delta k,K)$ is the Fourier transform of the 
core one-boson emission function, $S_c(x,p)$, 
and the subscripts $c$ and $h$ indicate the contributions from the 
core and the halo, respectively, see refs.~\cite{csorgo_96ch,csorgo_ch97}
for further details. If the core-halo model is applicable,
the intercept parameter $\lambda_*$ becomes a momentum-dependent measure
of the core/halo fraction as follows from eq.~(\ref{lambda1}).

If the $\eta'$ mass is decreased, a large fraction of 
the $\eta'$s will not be able to leave 
the hot and dense region through thermal
fluctuation since they need to compensate for the missing mass by large
momentum \cite{kapusta96etap,xnwang,shuryak}.  
These $\eta'$s will thus be trapped in the hot and dense region until it
disappears, after which their mass becomes normal again and 
as a consequence of this mechanism, 
they will have small $p_t$.  The $\eta'$s then decay to pions via 
\beq
\eta' \rightarrow \eta + \pi^+ + \pi^-  \rightarrow (\pi^0 + \pi^+ + \pi^-)
+ \pi^+ + \pi^-.
\eeq{etap_decay}
Assuming a symmetric decay configuration $(|p_t|_{\pi^+} 
\simeq |p_t|_{\pi^-} \simeq |p_t|_{\eta})$ 
and letting $m_{\eta'} = 958$ MeV, $m_{\eta} = 547$ MeV and 
$m_{\pi^+} = 140$ MeV, the average $p_t$ of the pions from the 
$\eta'$ decay is found to be $p_t \simeq 138$ MeV.
As the $\eta', \eta$ decays contribute to the halo due to their
large decay time ($1/{\Gamma_{\eta', \eta}} > 20$ fm/c),   
we expect a hole in the low $p_t$ region of the 
effective intercept parameter, 
$\lambda_* = [N_{core}({\bf p})/N_{total}({\bf p})]^2$, centered around
$p_t \simeq 138$ MeV.
If the masses 
of the $\omega$ and $\eta$ mesons also decrease in hot and dense matter,
this $\lambda_*(m_t)$ hole may even be deepened further, 
as discussed in ref.~\cite{vcsk}. 

\section{Numerical simulation}
In this section, we briefly review the essential steps of 
the numerical simulations and highlight some selected results 
following ref.~\cite{vcsk}.
In the numerical calculation of $\lambda_*$, we suppressed the rapidity
dependence by considering the central rapidity 
region, $( -0.2 < y < 0.2 )$ only.   
As a function of $m_t = \sqrt{p_t^2 + m^2}$,  $\lambda_*(m_t)$ is given by
eq.~(\ref{lambda1}), 
where the numerator represents the invariant $m_t$ distribution of $\pi^+$ 
emitted from the core and where the denominator represents the invariant $m_t$ 
distribution of the total number of $\pi^+$ emitted.  

To calculate the $\pi^+$ contribution from the halo region,  
the resonances 
$\omega$, $\eta'$, $\eta$ and $ K^0_S$ were given 
an $m_t$  according to the distribution~\cite{csorgo_96ch,3d} of 
\beq  
N(m_t) = C m_t^{\alpha} e^{-m_t/T_{eff}}, 
\eeq{mt_dist}
where C is a normalization constant, $\alpha = 1 - d/2$,
$d = 3$ is the dimension of the expansion,
and the mass-dependence of the slope parameter is\cite{3d,na44_teff}
\beq
T_{eff} = T_{fo} + m \langle u_t \rangle^2 . 
\eeq{Teff} 
with $T_{fo} = 140$ MeV being the freeze-out temperature 
and $\langle u_t \rangle =0.5 $ is the  average transverse flow velocity. 
This way, the long lived resonances were generated,
then they were decayed using
Jetset 7.4 \cite{jetset74_94}.    
The $m_t$ distribution of the core pions was 
also obtained from Eqs. (\ref{mt_dist}) and (\ref{Teff}).  
The contributions from the decay products of the core and the halo  
were then added together according to their respective fractions, 
as given in ref.~\cite{heiselberg_96},
allowing for the determination of $\lambda_*(m_t)$.    
The presence of the hot and dense region involves including  
an additional relative fraction of $\eta '$ with a medium 
modified $p_t$ spectrum. The $p_t$ spectrum of these $\eta '$ 
is obtained by assuming  energy conservation and zero longitudinal 
motion at the boundary between 
the two phases, 
\beq
m^{*2}_{\eta'} + {p_t}^{*2}_{\eta'} = m^{2}_{\eta'} + {p_t}^{2}_{\eta'}, 
\eeq{cons_mt}
where the ($*$) denotes the $\eta'$ in the hot dense region. 
The $p_t$ distribution then becomes a twofold distribution.  
The first part of the distribution is from the $\eta'$ which 
have $p_t^* \leq \sqrt{m^{2}_{\eta'} - m^{*2}_{\eta'}}$.   
These particles are given a $p_t = 0$.  
The second part of the distribution comes from the rest of the $\eta'$'s 
which have big enough $p_t$ to leave the hot and dense region.  These have the
same, flow-motivated $p_t$ distribution as the other produced resonances and 
were given a $p_t$ according to the $m_t$
distribution of eq.~(\ref{Teff}) with $d = 3$; the vacuum value
of the $\eta'$ mass, $m_{\eta'}$ was replaced by
the medium-modified $m_{\eta'}^*$ and the temperature of the 
hot and dense region was assumed to be $T' = 200$ MeV.
 
\begin{figure}[htb]
\begin{center}
\psfig{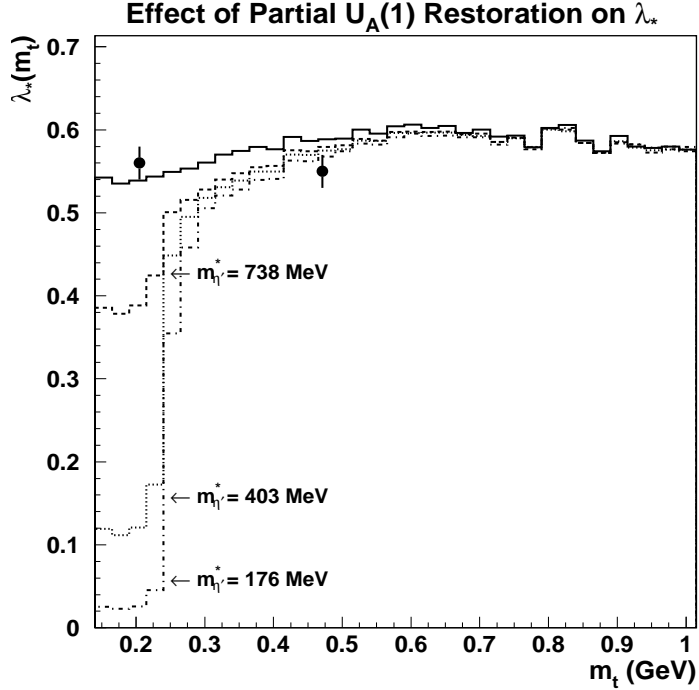}
\end{center}
\caption{
The solid line represents
$\lambda_*(m_t)$ assuming no partial $U_A(1)$ symmetry restoration
and normal $\eta'$ mass,
while the other lines represent the inclusion of hot and dense regions, 
where $T' = 200$ MeV and partial $U_A(1)$ symmetry restoration
results in a decreased mass of the $\eta'$ to 
$m^*_{\eta'} = 738$ MeV (dashed line), $m^*_{\eta'} = 403$ MeV (dotted line) 
and $m^*_{\eta'} = 176$ MeV (dot-dashed line).  All curves are calculated 
for $\langle u_t \rangle = 0.5$.
}
\end{figure}

Calculations of $\lambda_*(m_t)$ including the hot and dense regions 
are shown in Fig. 1. The abundances of long-lived resonances were estimated 
with the help of  the Fritiof model~\cite{fritiof_87}.
The two data-points shown on Fig. 1 are from S+Pb
reactions at 200 AGeV as measured by the NA44 collaboration~\cite{na44_mt}.
The lowering of the $\eta'$ mass and the partial chiral restoration result
in a deepening of the hole in the effective intercept parameter at low $m_t$.
This ``$\lambda_*$-hole" appears even for a 
modest enhancement of a factor of 3 in the $\eta'$ production, 
that corresponds to a slightly reduced effective mass, 
$m^*_{\eta'} =  738$ MeV. 
The effective masses $m^*_{\eta'} = 403$ MeV and $m^*_{\eta'} = 176$ MeV 
correspond to enhancement factors of 16 and 50, respectively.  
The onset of the full $U_A(1)$ symmetry
restoration in the $m_u = m_d = m_s = 0$ limiting case corresponds to
equal probability of the $\eta'$, $\eta$ and direct $\pi$ meson production,
in which case the intercept parameter $\lambda_*$ reaches its minimum value,
$\lambda_{U_A(1)} \simeq 0.02$, in the transverse mass region of 
$m_t \le 220$ MeV.  Thus a measurement of the intercept parameter
$\lambda_*$ in a large transverse mass interval may determine
whether a hole in the low $p_t$ region exists or not. If the
hole is present, its deepness characterizes the level of partial
$U_A(1)$ symmetry restoration in hot and dense matter. 
Full $U_A(1)$ restoration corresponds to the maximum size of the hole,
bottoming at a value of $\lambda_{U_A(1)}$. 
In this sense, the value of the 
$\Delta \lambda = \lambda_*(m_t) - \lambda_{U_A(1)} $ 
function in the $m_\pi \le m_t \le 220$ MeV region plays the role 
of an {\it experimentally measurable, effective
order parameter} of $U_A(1)$ symmetry restoration: its value 
is $ \Delta \lambda = 0$ for
the fully symmetric phase, while the inequality
$\Delta \lambda > 0$ is satisfied if the $U_A(1)$ symmetry
is not fully restored in hot and dense hadronic matter
produced in high energy heavy ion collisions. 

\section{Summary}
Partial $U_A(1)$
symmetry restoration in hot and dense hadronic matter
results in an observable hole for $m_t \le 220$ MeV region 
in   the shape of the $\lambda_*(m_t)$ function, that is
measurable by plotting the intercept
parameter of the two-pion Bose-Einstein correlation function versus
the mean transverse mass of the pair.
The  $\lambda_*$-hole signal of
partial $U_A(1)$ restoration cannot be faked in a conventional
thermalized hadron gas scenario, as it is not possible to create
significant fraction of the $\eta$ and $\eta'$
mesons with $p_t\simeq 0$ in such a case.
See refs.~\cite{vcsk,csorgo_sum} for 
further details and for a discussion of possible
coherence effects~\cite{csorgo_97prl}.
 
A qualitative analysis of NA44 S+Pb data suggests no visible sign of 
$U_A(1)$ restoration at SPS energies.  
The signal of partial $U_A(1)$ symmetry restoration 
should be searched for in Pb + Pb collisions at CERN SPS, 
and in forthcoming nuclear collisions 
at BNL RHIC and CERN LHC, by the experimental determination
of the $\lambda_*(m_t)$ function at $m_t <$ 220 MeV. 

\section*{Acknowledgments}
We thank M. Gyulassy, U. Heinz, J. Kapusta, L. McLerran,
X. N. Wang and U. Wiedemann for useful discussions.  
This work was supported by the OTKA Grants 
T024094, T026435, by the NWO - OTKA Grant N25487,
by the US- Hungarian Joint Fund MAKA 652/1998,
and by the Director, Office of Energy Research, 
Division of Nuclear Physics of the Office of High Energy 
and Nuclear Physics of the U.S. Department of Energy
under Contract No. DE-FG02-93ER40764.

\end{document}